\documentclass[aip,jap,onecolumn,superscriptaddress]{revtex4-1}
\usepackage{graphicx}

\begin{document}

\title{Carrier Density Dependence of the Magnetic Properties in Iron-doped Bi$_2$Se$_3$ Topological Insulator}

\author{H. Li}
\author{Y. R. Song}
\author{Meng-Yu Yao}
\author{Fengfeng Zhu}
\author{Canhua Liu}
\author{C. L. Gao}
\author{Jin-Feng Jia}
\author{Dong Qian}
\email{dqian@sjtu.edu.cn}
\author{X. Yao}
\affiliation{Key Laboratory of Artificial Structures and Quantum Control (Ministry of Education), Department of Physics, Shanghai Jiao Tong University, Shanghai 200240, China}
\author{Y. J. Shi}
\author{D.Wu}
\affiliation{The National Laboratory of Solid State Microstructures (NLSSM), Nanjing University, Nanjing, 210093, China}

\begin{abstract}

The electronic and magnetic properties of iron-doped topological insulator Bi$_{1.84-x}$Fe$_{0.16}$Ca$_x$Se$_3$ single crystals were studied. By co-doping Fe and Ca atoms, ferromagnetic bulk states with different carrier density (from n-type to p-type) were obtained. Effective magnetic moments for each Fe atom was estimated as small as about 0.07$\mu$$_{B}$. Magnetic and non-magnetic phase separation was observed in all samples. Our results suggest that the bulk ferromagnetism in Fe-doped Bi$_2$Se$_3$ is not intrinsic and regardless of carrier density.

\end{abstract}

\pacs{} \maketitle
\section{INTRODUCTION}

Distinguished from trivial insulators, topological insulators (TIs) - a new state of quantum matter - possessing gapless topological edge states protected by time reversal symmetry (TRS) have been attracting extensive studies recently\cite{HasanReview, ZhangReview, BernivegScience, KonigScience, FuPRB,Moore, HsiehNature, BiTeScience, ZhangNaturePhysics, XiaNaturePhysics, BiSeNature, SbTePRL, WangScience, YangPRL}. A number of materials have been proved or proposed to be TIs including Hg(Cd)Te quantum well\cite{KonigScience}, Bi$_{1-x}$Sb$_x$ alloys\cite{HsiehNature}, binary compounds (Bi$_{2}$Se$_{3}$, Bi$_{2}$Te$_{3}$, Sb$_{2}$Te$_{3}$)\cite{ZhangNaturePhysics,XiaNaturePhysics,BiTeScience,SbTePRL}, Half-Heusler compounds\cite{LinNatureMaterial,XiaoPRL}, ultra-thin Bi(111) bilayers\cite{Okada,YangPRL} and so on. Among these materials, Bi$_{2}$Se$_{3}$ is thought to be the most promising three dimensional TI for potential applications because of its simplest single helical Dirac cone surface states and the largest bulk gap (300meV)\cite{HsiehNature,ZhangNaturePhysics}. An energy gap can be opened at Dirac point to form massive Dirac Fermion by introducing TRS breaking field, such as magnetic field. Similar to graphene, opening an energy gap in the Dirac cone is very important for potential applications. In addition, the interplay between ferromagnetic states and topological states can cause many exotic phenomenon such as magnetic monopole, anomalous quantum Hall effect and the topological properties related Faraday and Kerr magneto-optical effects\cite{QiScience,YuScience,TsePRL}. Ferromagnetic states have been observed in 3d transitional magnetic elements (V, Cr, Mn, Fe) doped Bi$_{2}$Te$_{3}$, Bi$_{2}$Se$_{3}$, Sb$_{2}$Te$_{3}$ single crystals or films\cite{Choi,BTHor,Kulbachinskii,Dyck1,Dyck2,CrSTPRB,ZhouPRB,Haazen,LiAPL}. Massive Dirac Fermion will form when the magnetic moments have component perpendicular to the surface. In those magnetic TIs, the expected massive Dirac Fermion was only observed in Fe-doped and Mn-doped Bi$_{2}$Se$_{3}$ using angle-resolved photoemission spectroscopy (ARPES)\cite{ChenYL,YuNaturePhysics}. Though, the observed energy gap at Dirac point is very large ($\sim$60meV), the bulk magnetic properties of Fe$_{x}$Bi$_{2-x}$Se$_{3}$ are still controversial\cite{ChenYL,arxiv,CavaPRB}. At the same time, the Fermi level is not within the gap\cite{ChenYL} either. In the first report which claimed the observation of the massive Dirac Fermion\cite{ChenYL}, high Curie temperature (Tc) and small total magnetic moments were observed. Assuming doped Fe atoms have a valence of +3, only few Fe (x=0.002 - 0.005 ) atoms are estimated to enter the material matrix with a nominal concentration of x=0.1 to x=0.3. Intergrowth of Bi$_2$Se$_3$ and magnetic Fe$_x$Se$_y$ phase was also observed in some samples\cite{CavaPRB}. In another experiment\cite{arxiv}, Fe atoms were thought to completely enter the material matrix and make Fe-doped Bi$_{2}$Se$_{3}$ be intrinsic ferromagnetic semiconductor. Since Bi$_{2}$Se$_{3}$ is a semiconductor, in principle, intrinsic ferromagnetic TI would have similar magnetic behavior as the well-defined diluted magnetic semiconductor such as (Mn,Ga)As\cite{MnGaAs1,MnGaAs2}. If the ferromagnetic states in Fe-doped Bi$_{2}$Se$_{3}$ is intrinsic, the overall magnetic properties should strongly depend on the carrier density. On the other hand, in some systems, dopants with different valance can destroy the formation of ferromagnetic clusters\cite{NCrZnTe}. It is very important to carry out the carrier dependent experiments on Fe-doped Bi$_{2}$Se$_{3}$ systems for several reasons: i)To understand the origin of the bulk ferromagnetic states. ii)To control its magnetic properties. iii)To move the Fermi level into the gap near the Dirac point.

In this work, we carefully studied the magnetic properties of Fe-doped Bi$_{2}$Se$_{3}$ with Fe concentration of x=0.16 where the energy gap at Dirac point was nearly saturated\cite{ChenYL} as a function of carrier density. We succeeded in tuning the Fermi level to Dirac point with proper mount of Ca doping. For all Ca doped samples, ferromagnetic states were obtained. However, magnetic and non-magnetic phases separation existed in all samples. The magnetic properties could not be tuned systematically by carrier density. Regardless the carrier density, bulk ferromagnetism mainly came from extrinsic magnetic clusters.

\section{EXPERIMENT}
High quality single crystals of Bi$_{1.84-x}$Fe$_{0.16}$Ca$_x$Se$_3$ (x=0, 0.02, 0.04, 0.06) were grown by modified Bridgman method\cite{ChenYL}. High purity of Bi (99.999\%), Fe (99.999\%), Ca (99.5\%) and Se (99.999\%) powders were carefully mixed and sealed in evacuated quartz tubes. Excess amount of selenium was used to reduce the selenium vacancy. The tubes were firstly heated to 850$^\circ$C and kept for 24 hours, then slowly cooled to 550 $^\circ$C within 2 days, followed by 3 days annealing. The obtained crystals were well crystallized and could be easily cleaved perpendicular to the c-axis. The concentration of Fe and Ca dopants in the crystals were carefully determined by inductively coupled plasma (ICP) (iCAP THERMO 6000 Radial). The band structures were measured by ARPES with 10 eV photons at 77K in National Synchrotron Radiation Laboratory (NSRL, Hefei) using Scienta R4000 analyzer with base pressures better than 5$\times$10$^{-11}$ Torr. Hall measurements were carried out in the Magnetocryostat (Oxford). Magnetic measurements were performed by Physical Property Measurement System (PPMS) (Quantum Design) with magnetic field paralleled or perpendicular to the c-axis of the samples. ICP analysis shows that almost all of the nominal Fe atoms incorporated into the crystals. Though not all the Ca can enter the crystals, Ca concentration increases with x, approximately (Table-1). The maximum Ca entered into the crystals we obtained is about x$_{eff}$=0.018.

\section{RESULTS AND DISCUSSION}

As-grown Bi$_{2}$Se$_{3}$ crystals show electron doping (n-type) due to the intrinsic Se vacancies \cite{XiaNaturePhysics,41}. Altering the Bi:Se ratio in the nominal composition can reduce the defects to a certain extent, however, can not change the carrier type. If Fe atoms enter the Bi-Se matrix, they can occupy two possible positions in the crystal. The first possibility is that Fe atoms substitute Bi atoms with the same valence of +3, thus it will hardly change the carrier density\cite{ChenYL,YuScience,45}. This kind of substitution happens in other doped topological insulators\cite{GdBS,CrSTPRB,LiAPL}. Another possibility is Fe intercalation in the Van der Waals space between two quintuple layers. In this case, Fe atoms will supply additional electrons into Fermi sea\cite{AndrewNatPhy,PanPRL}. Figure 1 (a)-(d) present the low energy electronic structures of Bi$_{1.84-x}$Fe$_{0.16}$Ca$_x$Se$_3$ measured by ARPES at 77K. Linearly dispersive surface states are clearly observed in all samples. Similar to the reported Fe-doped samples\cite{ChenYL}, the Dirac point of surface states locates at $\sim$ 200 meV below the Fermi level without Ca doping. Fermi level touches the bottom of conduction bands. In Bi$_2$Se$_3$ single crystals, the Dirac point locates at about 200-350 meV below the Fermi level depending on the initial ratio of Bi and Se. Known from the ARPES results, Fe does not act as an electron donor in our samples, which indicates that there is very little Fe intercalation. Each Ca atom can substitute Bi atom in the form of Ca$^{2+}$ and supplies a hole into the Fermi sea\cite{BiTeScience}, so with the increase of Ca doping levels the Fermi level moves downward. Fermi level is tuned to the position that is very close the Dirac point(Fig. 1(d)) when the Ca nominal doping is x=0.06. Figure 1 (e) and (f) show the energy distribution curves (EDCs) near the Dirac points for x=0 and x=0.02, respectively. Blue curves are the EDCs at k=0. Spectral weight suppression ("gap"-like) at Dirac point were clearly observed on x=0 and x=0.02 samples. Seen from Fig. 1(e) and (f), though the energy resolution is not as good as the reference [30], two peaks in EDC (blue curve) were observed which is very similar to the previous report\cite{ChenYL}. For x=0.04 and 0.06 samples, because of the doping induced disorder, the spectra is not sharp enough to observe spectral weight suppression.  The gap we observed in x=0 and x=0.02 samples is about 50 to 90 meV depending on samples that is close to the gap value in reference [30]. However the "gap"-like feature is not the sole evidence of the formation of long-range intrinsic ferromagnetism. In fact, "gap"-like feature was observed in many topological insulators with magnetic or non-magnetic doping recently\cite{Yu}. We can not make conclusion on the magnetic states based on the ARPES results. ARPES measurement is surface sensitive, which can give the information of the carrier density near the surface region. The bulk carrier density as a function of Ca doping was further measured by Hall effects. Figure 1(g) presents the Hall resistivity as a function of applied magnetic field. Figure 1(h) shows the carrier density extracted from  Fig. 1(g) as a function of Ca concentration. The reported carrier density of Bi$_2$Se$_3$ varies from $\sim$10$^{19}$ cm$^{-3}$ to $\sim$10$^{17}$ cm$^{-3}$ obtained by different groups\cite{46,33,34} depending on the initial stoichiometry of Bi and Se and the preparing methods. In our parent compound Bi$_{1.84}$Fe$_{0.16}$Se$_3$, the measured carrier density is about 2.5$\times10^{18}cm^{-3}$ which is consistent with previous reports. Consistent with ARPES measurements, the incorporation of Ca into the crystals reduces the electron density and drives the carrier change from n-type to p-type. By co-doping Fe and Ca, we successfully tuned the Fermi level to the Dirac point. The successful tuning of the carrier concentration also allows us to explore the relation between magnetic properties and carrier density.

\begin{figure}
\includegraphics[width=8.5cm]{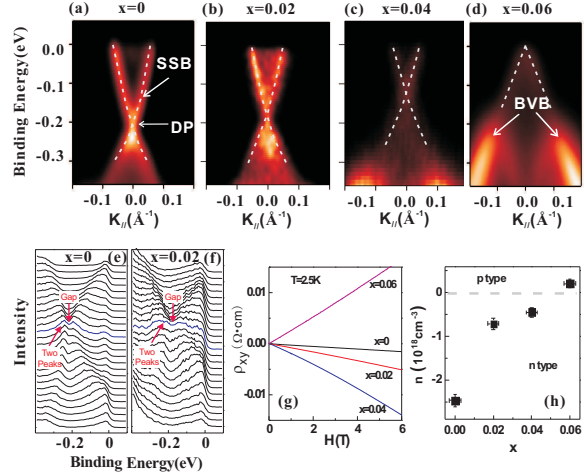}
\caption{Low energy ARPES spectra of Bi$_{1.84-x}$Fe$_{0.16}$Ca$_x$Se$_3$ with (a)x=0, (b)x=0.02, (c)x=0.04, (d)x=0.06. Fermi level moves downwards to the Dirac point with the increase of Ca dopants. (e)EDCs near the Dirac point for x=0 and (f)x=0.02. Blue curves are EDCs at k=0. (g) Hall resistivity as a function of applied magnetic filed. (h) Carrier density as a function of Ca concentration.}
\end{figure}

For all the samples, bulk ferromagnetism is observed at 2.1K. Figure 2 (a) and (b) show the magnetic moments (M) as a function of temperature (T) for different Ca concentrations with magnetic field perpendicular or parallel to the c-axis, respectively. Under the applied magnetic filed (1KOe), the amplitude of the magnetic signal varies for different Ca concentration, however, they show similar T-dependent behavior. Figure 2 (c) and (d) present the M-T curves after being normalized by the signals at the lowest temperature (2.1 K). In other magnetic topological insulators, like Cr-doped Sb$_2$Te$_3$\cite{CrSTPRB,LiAPL} and Mn-doped Bi$_2$Te$_3$\cite{BTHor}, only single magnetic transition point was observed. However, in Fe-doped Bi$_2$Se$_3$, there are three possible turn points in the M-T curves -- at about 10K, 100K to 200K and 250K to 300K, similar to previous results\cite{ChenYL, arxiv}. This implies the magnetic properties in this system are much more complicated.

\begin{figure}
\includegraphics[width=8.5cm]{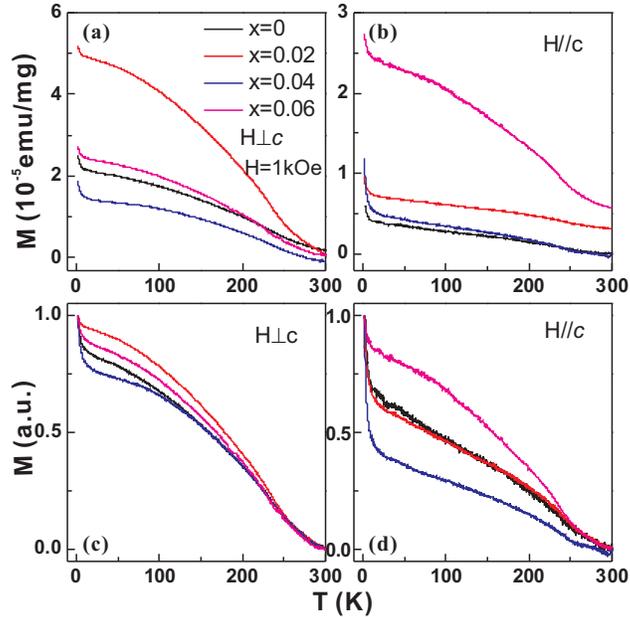}
\caption{Field cooled temperature dependent dc magnetism with an applied filed of 1kOe
for the Bi$_{1.84-x}$Fe$_{0.16}$Ca$_x$Se$_3$ crystals along (a) in-plane, (b)out-of-plane, (c)in-plane after normalization, (d) out-of-plane after normalization.}
\end{figure}

 \begin{figure}
\includegraphics[width=8.5cm]{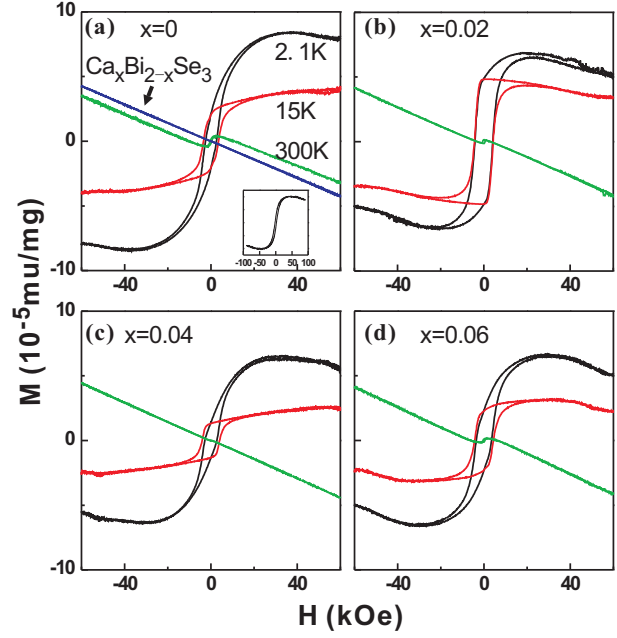}
\caption{The Magnetic hysteresis loops along the in-plane direction at different temperature for (a)x=0, (b)x=0.02 (c)x=0.04, (d)x=0.06}
\end{figure}

\begin{figure}
\includegraphics[width=8.5cm]{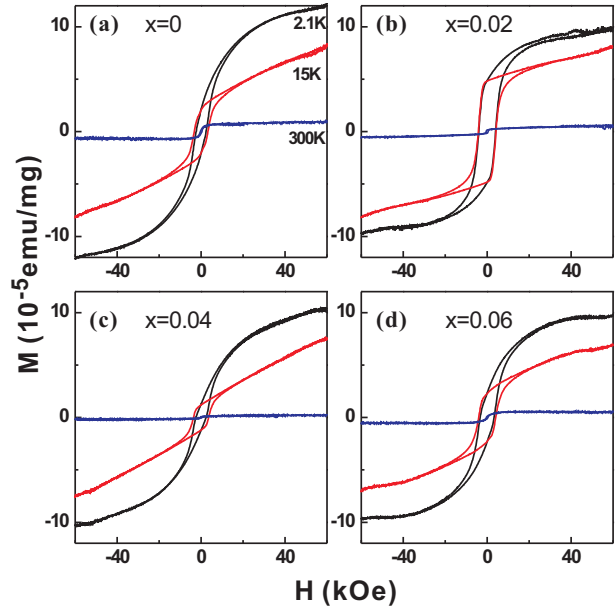}
\caption{The Magnetic hysteresis loops along the in-plane direction at different temperatures after subtracting the diamagnetic signals for (a)x=0, (b)x=0.02 (c)x=0.04, (d)x=0.06}
\end{figure}

In order to find the origin of the magnetic signals in Fe-doped Bi$_2$Se$_3$, we further systematically studied the magnetic hysteresis loops (M-H curves) along the in-plane direction as a function of carrier density. Figure 3 present M-H curves of samples with different Ca concentrations. The M-H curve of Bi$_2$Se$_3$ sample shows pure diamagnetic behavior, which does not show any noticeable change by small amount of Ca-doping. The blue curves present diamagnetic signals from the parent compounds without Fe. With Fe doping, there are clear hysteresis loops at 2.1K which indicates the existence of ferromagnetic state. Seen from Fig. 3(a), the magnetic signal is nearly saturated at about 40 KOe and a diamagnetic signal appears again at high magnetic field. Insert in Fig. 3(a) presents the magnetic signal up to 90 KOe. Diamagnetic signals are much clearer at high field. The estimated diamagnetic susceptibility is about 7.1$\times10^{-6}$emu/(g$\cdot$T) which is close to that of pure Bi$_2$Se$_3$. Similar diamagnetic signals are also obvious in Ca-doped samples (Fig. 3(b)-(d)). The existence of the diamagnetic signal implies that there is phase separation of non-magnetic Bi$_2$Se$_3$ or Ca-doped Bi$_2$Se$_3$ phase and magnetic phase. In Figure 4, We subtract the diamagnetic signals from the total signals to get the pure magnetic signals that related to the magnetic phase with Fe dopants. The overall M-H curves become simple. See from Fig. 5, at 2.1K, at zero field, the remanent magnetization is larger along the in-plane direction, which means the magnetic moments intend to stay in plane.

\begin{figure}
\includegraphics[width=8.5cm]{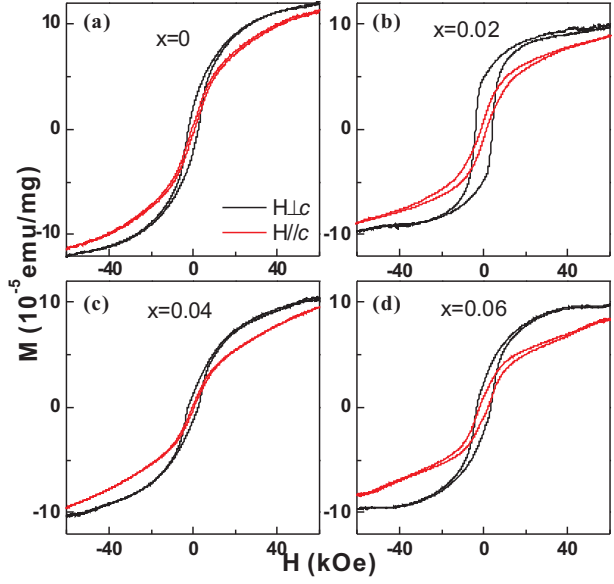}
\caption{Comparison of the magnetic hysteresis loops along the out-of-plane and in-plan directions after subtracting the diamagnetic signals for (a)x=0, (b)x=0.02 (c)x=0.04, (d)x=0.06}
\end{figure}

For all the samples, the pure magnetic signals are nearly saturated at 60 KOe. Using the signal measured at 60 KOe and ICP results, we calculated the average magnetic moments of per Fe atom as shown in Table 1. For all samples, the calculated magnetic moments per Fe atom are very close. It is about 0.07$\pm0.02 \mu_B$ which is much smaller than possible magnetic moments of Fe$^{3+}$ ion. Such small magnetic moment can not been explained by the scenario of Fe substitution of Bi. Actually, first principle calculation\cite{YuScience} assuming Fe substitution of Bi gave high spin states of Fe atoms. Since the magnetic properties of Fe-doped Bi$_2$Se$_3$ do not obviously depend on the carrier density, we conclude that the most part of the magnetic signals is extrinsic. They are not diluted ferromagnetic semiconductors. We can exclude the formation of Fe inclusions. If the ferromagnetism comes from Fe inclusions, we should get large average magnetic moments per Fe atoms(2.2$\mu_B$/atom for bulk Fe). On the other hand, Fe and Se can form complicated compounds Fe$_x$Se$_y$\cite{IEEE,FeSeMag} which present ferrimagnetic and/or antiferromagnetic behaviors. According to previous report on bulk Fe$_x$Se$_y$ compounds\cite{FeSeMag}, the value of magnetic moment of Fe atom in Fe$_x$Se$_y$ varies depending on the exactly ration of Fe and Se. The reported minimum value is from 0.2 to 0.4 $\mu_B$ and has the same order of magnitude as our results. So we think that the magnetic phase in our Fe-doped Bi$_2$Se$_3$ samples could be some forms of Fe$_x$Se$_y$ nanocrystals. Two possible explanation on the very small magnetic momentum. First, those nanocrystals are ferrimagnetic with low spin states. Second, they can be in antiferromagnetic states with uncompensated surface spins that gives the small ferromagnetic signals. Since our finding is based on bulk measurement, we can not exclude some intrinsic ferromagnetism on the surface which may open a real gap in ARPES spectra.

\begin{table*}
\caption{Concentrations of doped Ca and Fe and the magnetic moments for different magnetic dopants}
\begin{ruledtabular}
\begin{tabular}{cccccc}
x&Fe&Ca&n(cm$^{-3})$&M$_{6T}(\mu_B/Fe_{atom})$ \\ \hline
0&0.16$\pm$0.02 &0&-2.47$\times10^{18}$ &0.08$\pm$0.01\\
0.02&0.19$\pm$0.02 &0.009$\pm$0.005&-7.04$\times10^{17}$ &0.07$\pm$0.01\\
0.04&0.16$\pm$0.02 &0.012$\pm$0.005&-4.46$\times10^{17}$ &0.06$\pm$0.01\\
0.06&0.14$\pm$0.02 &0.018$\pm$0.005&2.14$\times10^{17}$ &0.06$\pm$0.01\\
\end{tabular}
\end{ruledtabular}
\end{table*}

\section{CONCLUSIONS}

By co-doping Ca and Fe, we succeeded in tuning the carrier density from n-type to p-type in ferromagnetic Bi$_{1.84-x}$Fe$_{0.16}$Ca$_x$Se$_3$ samples. We studied the evolution of the magnetic properties as a function of carrier density. Contrast to well-defined ferromagnetic semiconductor, the magnetic properties in Fe-doped Bi$_2$Se$_3$ do not have obvious dependence on the carrier density. Our results suggest that the bulk ferromagnetic behaviors of Fe-doped Bi$_2$Se$_3$ are due to Fe$_x$Se$_y$ nanoclusters in the crystals. In principle, the nature of the gap observed near Dirac point could be solved by determining the spin polarization using spin-resolved ARPES or spin-polarized scanning tunnelling spectroscopy in the future.

\section{ACKNOWLEDGMENTS}
We thank Z. Sun and G. B. Zhang from National Synchrotron Radiation Laboratory of China for help on ARPES measurement. This work is supported by National Basic Research Programm of China (Grant No. 2012CB927401, 2011CB921902, 2013CB921902, 2011CB922200),NSFC (Grant No. 91021002, 11174199, 11134008, 11274228), Shanghai Committee of Science and Technology, China (No. 11JC1405000, 11PJ1405200, 12JC1405300), Shanghai Municipal Education Commission (11ZZ17) and SRF for ROCS, SEM. D.Q. acknowledges additional support from Programm for Professor of Special Appointment (Eastern
Scholar) at Shanghai Institutions of Higher Learning.

\end{document}